\documentclass[aps, prl, a4paper, showpacs, twocolumn, superscriptaddress,10pt, nofootinbib]{revtex4-1}

\usepackage{textcomp,enumerate, cancel}   
              			
\usepackage[margin=0.8in]{geometry}

\usepackage{color} 
\usepackage{amsmath,amssymb,amsthm,bbm,bm,nicefrac}   				
\usepackage{thmtools} 
\usepackage{graphicx,epsfig}                                                               		
\usepackage{multirow}
\usepackage{subfigure}
\usepackage{zx-calculus}
\usepackage{slashed}
\usepackage{orcidlink}

\newcommand{\ssection}[1]{\emph{#1.---}}

\usepackage{soul}

\usepackage{stmaryrd}
\usepackage{tikz}
\usepackage{mleftright}
\mleftright
\usetikzlibrary{matrix}
\usepackage{tikz}
\usetikzlibrary{shapes,arrows,chains, decorations.pathmorphing, decorations.pathreplacing}
\tikzset{quantum/.style={decorate, decoration=snake}}

\newcommand{\ket}[1]{\left|#1\right\rangle}

\newcommand{\ketbra}[2]{\left|#1\rangle\langle#2\right|}

\newcommand{\abs}[1]{\lvert #1\rvert}

\newcommand{\norm}[1]{\left\| #1\right\|}

\newcommand{\tr}[1]{\mathrm{Tr}\left[#1\right]}
\newcommand{\ptr}[2]{\mathrm{Tr}_{#1}\big[#2\big]}

\NewDocumentCommand{\pr}{som}{\Pr\IfBooleanT#1{\IfValueT{#2}{_{\substack{#2}}}}\left[\,#3\IfBooleanF#1{\IfValueT{#2}{\bigm\vert #2}}\,\right]}
\NewDocumentCommand{\esp}{som}{\IfBooleanTF#1{\IfValueTF{#2}{\underset{\subalign{#2}}{\mathbb{E}}}{\mathbb{E}}}{\mathbb{E}}\left[\,#3\IfBooleanF#1{\IfValueT{#2}{\bigm\vert #2}}\,\right]}
\newcommand{\interacts}{\ensuremath{\kern-2pt\leftrightsquigarrow\kern-2pt}}
\newcommand{\Alice}{\ensuremath{\mathsf{A}}}
\newcommand{\Bob}{\ensuremath{\mathsf{B}}}
\newcommand{\Verif}{\ensuremath{\mathsf{V}}}
\newcommand*{\eps}[0]{\epsilon}
\NewDocumentCommand{\game}{m}{\ensuremath{\mathsf{Game}_{#1}}}
\NewDocumentCommand{\accept}{}{\ensuremath{\mathsf{Accept}}}
\NewDocumentCommand{\reject}{}{\ensuremath{\mathsf{Reject}}}
\NewDocumentCommand{\xor}{}{\oplus}
\usepackage{mathtools}
\newcommand*{\eqdef}{\coloneqq}
\newcommand*{\eqAboveWidth}[2][=]{\mathrel{\stackrel{\substack{#2}}{#1}}}

\newcommand*{\eqeq}[2][=]{\eqAboveWidth[#1]{(\ref{#2})}}
\newcommand*{\leqref}[1]{\eqeq[\leq]{#1}}

\makeatletter
\newcommand{\sample}{\mathrel{\mathpalette\sample@\relax}}
\newcommand{\sample@}[2]{%
  \ooalign{%
    \hspace{\stretch{2}}\raisebox{0.7\height}{$\m@th\demotestyle{#1}\mathdollar$}\hspace{\stretch{1}}\cr
    $\m@th#1\leftarrow$\cr
  }%
}
\newcommand{\demotestyle}[1]{%
  \ifx#1\displaystyle\scriptstyle\else
  \ifx#1\textstyle\scriptstyle\else
  \scriptscriptstyle\fi\fi
}
\makeatother

\newcommand*{\cA}{\mathcal{A}}

\def\gets{\rightarrow}

\usepackage{slashed}
\usepackage{enumitem}

\theoremstyle{plain}

\newtheorem{theorem}{Theorem}

\newtheorem{remark}[theorem]{Remark}
\newtheorem{prop}[theorem]{Proposition}
\newtheorem{definition}[theorem]{Definition}

\newtheorem{lemma}[theorem]{Lemma}

\usepackage{MnSymbol}

\definecolor{secondaryColor}{HTML}{5869bc}

\usepackage{hyperref}
\hypersetup{colorlinks, allcolors=secondaryColor}
\usepackage{url}
\usepackage{cleveref}
\crefname{figure}{Fig.}{Fig.}
\crefname{theorem}{Theorem}{Theorems}
\crefname{prop}{Proposition}{Propositions}
\crefname{observation}{Observation}{Observations}
\crefname{definition}{Definition}{Definitions}
\crefname{ex}{Example}{Examples}
\crefname{lemma}{Lemma}{Lemmas}
\crefname{corollary}{Corollary}{Corollaries}
\crefname{result}{Result}{Results}
\crefname{attack}{Attack}{Atacks}

\begin{document}
 \title{A quantum cloning game with applications to quantum position verification}
\author{Lloren\c c Escol\`{a}-Farr\`{a}s, \orcidlink{0000-0001-6194-0491}}
\affiliation{QuSoft \& CWI Amsterdam, The Netherlands}
\affiliation{QuSoft \& Informatics Institute, University of Amsterdam, The Netherlands }

\author{Léo Colisson Palais\,\orcidlink{0000-0001-8963-4656}}
\affiliation{QuSoft \& CWI Amsterdam, The Netherlands}
\affiliation{Laboratoire Jean Kuntzmann, Université Grenoble Alpes, France}

\author{Florian Speelman, \orcidlink{0000-0003-3792-9908}}
\affiliation{QuSoft \& CWI Amsterdam, The Netherlands}
\affiliation{QuSoft \& Informatics Institute, University of Amsterdam, The Netherlands }

\begin{abstract}
We introduce a quantum cloning game in which $k$ separate collaborative parties receive a classical input, determining which of them has to share a maximally entangled state with an additional party (referee).  We provide the optimal winning probability of such a game for every number of parties $k$, and show that it decays exponentially when the game is played $n$ times in parallel. These results have applications to quantum cryptography, in particular in the topic of quantum position verification, where we show security of the routing protocol (played in parallel), and a variant of it, in the random oracle model. 
\end{abstract}

\maketitle

\ssection{Introduction} 
Non-local correlations have extensively been studied in the field of quantum information theory, see e.g.~\cite{Bell_non_locality_report}. Bell~\cite{Bell_1964} originally showed that distant parties sharing quantum resources can reproduce correlations that could never be attained by any classical theory. Often, non-local correlations are studied as \emph{non-local games}, which provide an operational framework for understanding them.  These games are interesting per se from a fundamental point of view, since they give rise to understanding the underlying essence of nature, but they additionally lead to applications such as secure key distribution \cite{acin_device-independent_2007}, 
certified randomness \cite{pironio_random_2010}, 
reduced communication complexity \cite{buhrman_nonlocality_2010},
self-testing \cite{mayers_self_2004,supic_self-testing_2019},
and computation \cite{anders_computational_2009}.

A vast literature in non-local games covers the scenario where a classical referee sends questions to non-communicating collaborative parties, and their task is to produce answers according to a certain publicly-known predicate, where the questions and answers are all \emph{classical}. The best-known non-local game is the CHSH game~\cite{CHSHArticle}.  Non-locality has also been investigated in terms of supersets of non-local games, called \emph{monogamy-of-entanglement (MoE) games}~\cite{TomamichelMonogamyGame2013}, where a quantum referee sends the same classical question to the players and the parties have to guess the (classical) outcome of a referee's quantum measurement (depending on the question). MoE games have been used to provide security proofs for the quantum cryptographic primitives device-independent quantum key distribution~\cite{BB84} and quantum position verification~\cite{OriginalQPV_Kent2011}. Such games were later generalized under the name of \emph{extended non-local games}~\cite{Extended_non-local_games_andMoE_games}. 

Here, we introduce the concept of the \emph{quantum cloning game}, played by $k$ distant parties and a quantum referee. The referee publically announces a party, i.e., sends the same classical question to all the players, and the chosen party has to end up with the maximally entangled (EPR) state with the referee. At the beginning of the game, the players are allowed to share any quantum state with the referee. In this work, we show that the optimal winning probability for players using any quantum resources is given by $\frac{1}{2}+\frac{1}{2k}$, converging to $\frac{1}{2}$ for a large number of players. We analyze the game when it is played $n$ times in parallel, showing an exponential decay in $n$ of the optimal winning probability. Additionally, the quantum cloning game can be generalized to any arbitrary quantum state instead of an EPR state, and we provide its optimal winning probability.

We show that these results have applications in quantum position verification (QPV), 
which is a cryptographic primitive consisting of verifying the location of an untrusted party. Securely implementing this primitive is unachievable using only classical information, because a general attack exists even when using computational assumptions~\cite{OriginalPositionBasedCryptChandran2009}.
Due to the no-cloning theorem~\cite{Wootters1982NoCloning} the general classical attack does not apply if quantum information is used instead \cite{OriginalQPV_Kent2011,Malaney_QLP}, however, a general quantum attack exists which requires exponential entanglement~\cite{Buhrman_2014,Beigi_2011}. This means that hope for protocols secure against reasonable amounts of entanglement is alive, and indeed there has been much analysis on attacks on specific protocols~\cite{PatentKentANdOthers,OriginalQPV_Kent2011,Lau_2011,https://doi.org/10.48550/arxiv.1504.07171,Chakraborty_2015,speelman2016instantaneous,dolev2019constraining,dolev2022non,gonzales2019bounds,cree2022code}, and security analysis under extra assumptions ~\cite{liu2021beating,gao2016quantum}, such as the random oracle model~\cite{Unruh_2014_QPV_random_oracle}. A generic 1-dimensional (the main ideas generalize to higher dimensions) QPV protocol is described in the following way: two verifiers \textsf{V}$_0$ and \textsf{V}$_1$, placed on the left and right of an untrusted prover \textsf{P}, supposedly at the position $pos$, send quantum and classical messages to \textsf{P} at the speed of light, and he has to pass a challenge and reply correctly to them at the speed of light as well, if so, the verifiers \emph{accept},  and if any of them receives a wrong answer or the timing does not correspond with the time it would have taken for light to travel back from the honest prover, the verifiers \emph{reject}. The time consumed by the prover to perform the challenge is assumed to be negligible, and the verifiers are assumed to have perfectly-synchronized clocks. 

In this work, we consider the \emph{routing} QPV protocol~\cite{OriginalQPV_Kent2011}, which has an appealing simple form: the prover has to return a received qubit to one of the verifiers, where the choice of verifier is a function of the classical information sent by the verifiers~
\cite{OriginalQPV_Kent2011}. Besides the theoretical interest of this protocol, it is also an appealing candidate for free-space quantum position verification, when the quantum messages can travel with the vacuum speed of light, since the hardware of the prover could hypothetically be as simple as a mirror or an optical switch.
Despite theoretical work on this protocol~\cite{,Buhrman_2013,cree2022code,bluhm2022single,Allerstorfer2024relatingnonlocal,asadi2024lineargateboundsnatural}, there were gaps left in our understanding relative to measurement-based QPV protocol variants: namely (i) the security of parallel repetition of this protocol against unentangled attackers, and (ii) its security in the random-oracle model against arbitrary adversaries. As an application of the quantum cloning game, we show the security of the routing protocol in these scenarios.

In this paper, we state the main results and brief intuitions behind the proofs. See supplemental material (SM) for comprehensive details, extensive results, and complete proofs. In all the figures, classical and quantum information are represented by straight and undulated lines, respectively.

\ssection{Preliminaries} 
For $k\in\mathbb N$, we will denote $[k]:=\{0,\ldots,k-1\}$. Let $\mathcal{H}$, $\mathcal{H'}$ be finite-dimensional Hilbert spaces, we denote by $\mathcal{B}(\mathcal{H},\mathcal{H'})$ the set of bounded operators from $\mathcal{H}$ to $\mathcal{H'}$ and $\mathcal{B}(\mathcal{H})=\mathcal{B}(\mathcal{H},\mathcal{H})$. 
Denote by $\mathcal{S}(\mathcal{H})$ the set of quantum states on $\mathcal{H}$,~i.e.\ ${\mathcal{S}(\mathcal{H})=\{\rho\in\mathcal{B}(\mathcal{H})\mid \rho\geq0, \tr{\rho}=1)\}}$. A pure state will be denoted by a ket $\ket{\psi}\in\mathcal{H}$. The maximally entangled state or EPR pair is ${\ket{\Phi^+}=\frac{1}{\sqrt{2}}(\ket{00}+\ket{11})}$. We denote the identity matrix by $\mathbb I$. For $M\in\mathcal{B}(\mathcal{H})$, we denote its Schatten $\infty$-norm by $\norm{M}$. We will use the notation  $1,\ldots,\slashed{i},\ldots, k-1$ to denote $1,\ldots, i-1,i+1,\ldots, k-1$. 

\begin{figure}[h]
\centering
\includegraphics[width=75mm]{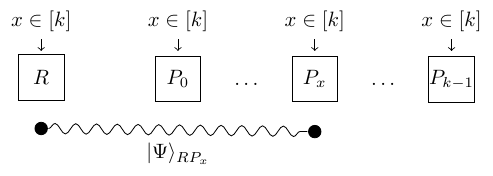}
\caption{Schematic representation of the $k$-party quantum cloning game, where $\ket{\Psi}_{RP_x}=\ket{\Phi^+}_{RP_x}$. If $\ket{\Psi}_{RP_x}$ is arbitrary, this represents  a $\Psi$-QCG$_k$.}
\label{Fig_cloning_Psi_k_parties}
\end{figure}

\ssection{$k$-party quantum cloning game} In the following definition, we introduce the quantum cloning game. 
\begin{definition}\label{def:kpartyQCG} The $k$-party \emph{quantum cloning game}, shortly denoted by \emph{QCG}$_k$, consists of a referee $R$ with associated Hilbert space $\mathcal{H}_R=\mathbb{C}^2$ and $k$ collaborative distant parties (players) $P_0,\ldots,P_{k-1}$. Before the game starts, the parties prepare a joint quantum state of arbitrary dimension between themselves and the referee. During the game, the referee sends $x\in[k]$, drawn uniformly at random, to all the collaborative parties. The players win the game if and only if the party $P_x$ (holding a qubit register $P_x$) ends up sharing the maximally entangled state with the referee, i.e.~if a projection onto $\ket{\Phi^+}_{RP_x}$ yields the correct outcome.
\end{definition}
See \cref{Fig_cloning_Psi_k_parties} for a schematic representation of the ${\text{QCG}_k}$. Intuitively, in such a game, the referee publically announces which party has to create an entangled state with herself. 

A strategy ${S}$ for the QCG$_k$ is described by a quantum state
$\rho\in \mathcal{S}(\mathcal{H}_R\otimes\mathcal{H}_{P_0E_0}\otimes\cdots\otimes\mathcal{H}_{P_{k-1}E_{k-1}})$, where, for $i\in[k]$, registers $P_i$ are of the same dimension as $\mathcal{H}_R$ and $E_i$ are auxiliary systems of arbitrary dimension that each party possess, and completely positive trace-preserving (CPTP) maps $\{\mathcal{E}^x_{P_iE_i\rightarrow P_i}\}_{x}$, where the subscript $P_iE_i\rightarrow P_i$ indicates that the map has input and output registers $P_iE_i$ and $P_i$, respectively, i.e. $\mathcal{E}^x_{P_iE_i\rightarrow P_i}:\mathcal{B}(\mathcal H_{P_iE_i})\rightarrow \mathcal{B}(\mathcal H_{P_i})$.  The winning probability of such a game, given the strategy $S$, is provided by
\begin{equation}
\begin{split}
    &\omega(\text{QCG}_k,S)=\frac{1}{k}\sum_{x\in [k]}\text{Tr}\bigg[\ketbra{\Phi^+}{\Phi^+}_{VP_x}\\&
    \ptr{P_0\ldots \slashed{P}_{x} \ldots P_{k-1}}{\mathbb I_{R}\bigotimes_{i\in[k]} \mathcal{E}^x_{P_iE_i\rightarrow P_i}(\rho)}\bigg].
\end{split}
\end{equation}
The optimal winning probability of such games is given by 
\begin{equation}
    \omega^*(\text{QCG}_k)=\sup_{S}\omega(\text{QCG}_k,S),
\end{equation}
where the supremum is taken over all the possible strategies over all possible Hilbert spaces. The following theorem gives the optimal winning probability of this game for every number of parties $k$. 

\begin{theorem}\label{thm:w_k_parties_EPR}
    For every $k\in\mathbb N$, the optimal winning probability of the $\emph{\text{QCG}}_k$ is given by
    \begin{equation}
        \omega^*(\emph{\text{QCG}}_k)=\frac{1}{2}+\frac{1}{2k}.
    \end{equation}
\end{theorem}
Intuitively, this game cannot be perfectly won since, otherwise, it would be possible to have maximal entanglement between the referee and each of the parties, and this is not possible since entanglement is \emph{monogamous}~\cite{Coffman_2000}. The proof of \cref{thm:w_k_parties_EPR} can be found in the SM, where the key part is to show that the optimal winning probability is attainable by the actions of the players being \emph{independent} of $x$, intuitively, each party acts as if they were chosen to reproduce the maximally entangled state with the referee. In addition, in the proof, we show that the optimal value can be attained by preparing an initial state $\rho$ where, locally, each of the parties holds a qubit and no further actions taken by the players, i.e.\ their local actions are described by the identity channel  $(\mathbb I_{P_i})$. We then specify a strategy by providing a quantum state, since any local actions are independent of $x$, they can be absorbed in the quantum state.   More precisely, the optimal winning probability for the $\text{QCG}_k$ can be attained by the strategy given by the (pure) quantum state 
\begin{equation}\label{eq:optimal_strategy}
    \ket{\varphi}=\sqrt{\frac{2}{k(k+1)}}\sum_{x\in[k]}\ket{\Phi^+}_{RP_x}\ket{0}_{P_0\ldots \slashed{P}_{x} \ldots P_{k-1}}.
\end{equation}
Note that other natural multi-party entangled states that have been widely studied in the literature, such as the GHZ state~(\cite{Greenberger1989}) $\ket{GHZ}=\frac{1}{\sqrt{2}}(\ket{000}+\ket{111})$  and the W state~(\cite{PhysRevA.62.062314}) $\ket W=\frac{1}{\sqrt{3}}(\ket{001}+\ket{010}+\ket{100})$, and their respective generalizations to arbitrary dimensions, as well as the strategy of `guessing' which party has to reproduce the quantum state, e.g.\ guessing $x=0$, given by preparing the state $\ket{\Phi^+}_{VP_0}\ket{0}_{P_1}\ldots\ket{0}_{k-1}$, provide significantly suboptimal winning probabilities. For 2-players, $ \omega^*(\text{QCG}_2)=\frac{3}{4}$, and 
\begin{equation}
    \omega^*(\text{QCG}_k)\xrightarrow[]{k\rightarrow\infty} \frac{1}{2},
\end{equation}
which converges to the value attained by the strategy given by preparing the state ${ \ket{0}_R\ket{0}_{P_0}\ldots\ket{0}_{P_{k-1}}}$, showing that when $k$ increases even unentangled states allow for a near-optimal winning probability.

\ssection{Quantum cloning game with any target state} The concept of QCG$_k$ can be generalized to the case where, instead of the parties having to reproduce EPR pairs with the referee, the state that has to be reproduced is an arbitrary-fixed state, i.e.\ the referee's Hilbert space $\mathcal{H}_R$ is now of arbitrary dimension, and on input $x$ the party $P_x$ has to generate a given state $\ket{\Psi}_{RP_x}$. Here, the dimension of the registers $P_i$ is the same for all $i\in[k]$.  We will refer to such a game as a $k$-party \emph{quantum cloning game with target state } $\ket{\Psi}$, in short denoted by $\Psi$-QCG$_k$, see \cref{Fig_cloning_Psi_k_parties}.  Notice that this game becomes trivial if the target state $\ket{\Psi}_{RP}$ is a tensor product state. In the following theorem, we provide the optimal winning probability for any $\Psi$-QCG$_k$ for every number of parties $k$ and for any target state $\ket{\Psi}$.  
\begin{theorem}\label{thm:bound_k_parties} The optimal winning probability for every $\Psi$-\normalfont{QCG}$_k$ is given by
\begin{equation*}
    \omega^*(\Psi\text{\normalfont{-QCG}}_k)=\frac{1}{k}\norm{\sum_{x\in [k]}\ketbra{\Psi}{\Psi}_{VP_x}\otimes \mathbb{I}_{P_0\ldots \slashed{P}_{x} \ldots P_{k-1}}}.
\end{equation*}
\end{theorem}
See SM for the proof of \cref{thm:bound_k_parties}. Along the lines of the proof of \cref{thm:w_k_parties_EPR}, the key idea relies on showing that the optimal winning probability can be attained by the actions of the players being independent on $x$. 

\begin{figure}[h]
\centering
\includegraphics[width=70mm]{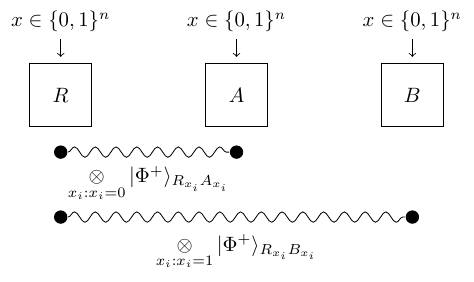}
\caption{Schematic representation of the $n$-fold parallel repetition of the $2$-party quantum cloning game. }
\label{Fig_parallel_EPR}
\end{figure}

\ssection{Parallel repetition of \normalfont{QCG}$_k$} A case of particular interest is given when $\text{QCG}_k$ is played $n$ times in parallel, denoted by $\text{QCG}_k^{\times n}$. Specifically, we will analyze $\text{QCG}_2$ where now the two collaborative parties, who we rename as Alice and Bob, will receive $x=x_0\ldots x_{n-1}\in\{0,1\}^n$. We denote by $R_0\ldots R_{n-1}$, $A_0\ldots A_{n-1}$ and $B_0\ldots B_{n-1}$ the final (qubit) registers of the referee, Alice and Bob, respectively. The players win if at the end of the game Alice is able to create the maximally entangled state with the referee in all her registers such that $x_i=0$, and analogously for Bob in all his registers such that $x_i=1$. See \cref{Fig_parallel_EPR} for a schematic representation. 
Similarly as before, at the beginning of the game the three parties are allowed to share any arbitrary quantum state and, upon receiving the classical information, Alice and Bob can apply CPTP maps $\{\mathcal{E}^x_{A_0\ldots A_{n-1}E_A\rightarrow A_0\ldots A_{n-1}}\}_x$ and $\{\mathcal{E}^x_{B_0\ldots B_{n-1}E_B\rightarrow B_0\ldots B_{n-1}}\}_x$, where $E_A$ and $E_B$ are arbitrary auxiliary systems that Alice and Bob possess, respectively. 

In the following theorem, we state that the optimal winning probability decays exponentially with the number of parallel repetitions $n$. 

\begin{theorem}\label{thm:optimal2QCGparallel} The optimal winning probability for $n$ parallel repetitions of the $\normalfont{QCG}_2$ is such that 
\begin{equation}\label{eq:parallel_rep_2_parties}
    \left(\frac{3}{4}\right)^n \leq\omega^*(\emph{\text{QCG}}_2^{\times n})\leq\left(\frac{1}{2}+\frac{1}{2\sqrt{2}}\right)^n.
\end{equation}
\end{theorem}

See SM for the proof of \cref{thm:optimal2QCGparallel}. The key idea of the proof relies on combining ideas used in the proof of \cref{thm:bound_k_parties} together with Proposition~4.3 in \cite{schaffner2007cryptographyboundedquantumstoragemodel}, which was also used in~\cite{TomamichelMonogamyGame2013} to prove parallel repetition for monogamy-of-entanglement games (for completeness, see ~\cref{lem:sum_projectors} in the SM). 

\begin{figure}[h]
\centering
\includegraphics[width=82mm]{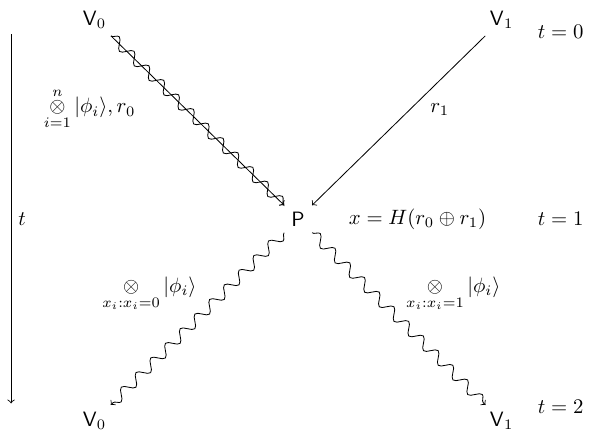}
\caption{Schematic representation of the $(H,n)$-routing QPV protocol. If $r_0$ is an empty bit string, and $x=r_1$, this figure represents the $n$-parallel repetition of the routing QPV protocol. The time arrow is represented by $t$.}
\label{Fig_H_routing}
\end{figure}

\ssection{Application to QPV in the No-PE model}  In this section, we analyze the security of the \emph{routing} QPV protocol, originally introduced in~\cite{OriginalQPV_Kent2011}. A round of this protocol, see \cref{Fig_H_routing} for a schematic representation, is described as follows: 
\begin{enumerate} [align=left,leftmargin=*,itemsep=0pt,parsep=0pt,topsep=1pt]
\item The verifier \textsf{V}$_0$ selects a qubit ${\ket{\phi}\in\{\ket0,\ket1,\ket+,\ket-\}}$, and the verifier \textsf{V}$_1$ selects $x\in\{0,1\}$, both picked uniformly at random. They send $\ket\phi$ and $x$ (at time $t=0$) so that they arrive at the same time ($t=1$) at $pos$. 
\item Upon receiving the information sent by  \textsf{V}$_0$ and  \textsf{V}$_1$, the prover sends the qubit $\ket\phi$ to the verifier \textsf{V}$_x$. 
\item If  $\ket\phi$ arrives at the time consistent with $pos$ ($t=2)$, and a projective measurement performed by \textsf{V}$_x$ on the state sent by \textsf{V}$_0$ leads to the correct outcome, the verifiers accept. Otherwise, they reject. 
\end{enumerate}

\begin{figure}
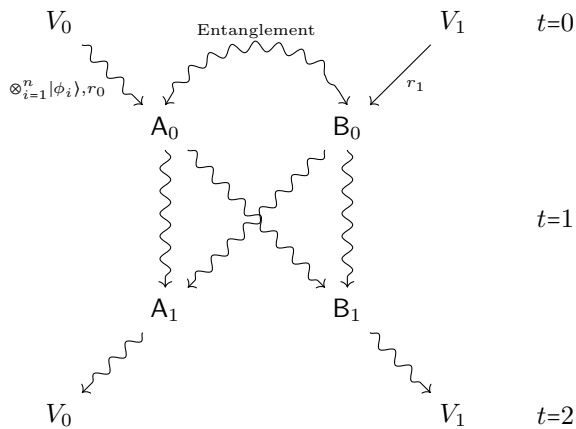
\centering
  \begin{ZX}[column sep=8mm,row sep=8mm,myar/.style={->,decorate, decoration={amplitude=.7mm,segment length=3mm,post length=1mm,snake,#1}},
    /utils/exec={%
      \def\aliceArrows{%
        \ar[myar,rdd]
        \ar[r,o',start anchor=north,end anchor=north,myar={pre length=2mm,post length=2mm},<->, "\text{Entanglement}" {yshift=1mm}]
        \ar[dd,myar]
      }%
    }]
    \zxElt{V_0} \ar[rd,myar,"{\otimes_{i=1}^n \ket{\phi_i}, r_0}" {swap}] &                               & [1cm]                                      & \zxElt{V_1} \ar[dl, ->, "r_1"] & t=0 \\
                                                                          & \zxElt{\Alice_0} \aliceArrows & \zxElt{\Bob_0} \ar[myar,ddl]  \ar[dd,myar] &                           &     \\
                                                                          &                               &                                            &                           & t=1 \\
                                                                          & \zxElt{\Alice_1} \ar[myar,dl] & \zxElt{\Bob_1} \ar[myar,rd]                &                                 \\
    \zxElt{V_0}                                                           &                               &                                            & \zxElt{V_1}               & t=2 \\[-6mm]
  \end{ZX}
  \caption{Schematic representation of a generic attack to the $(H,n)$-routing protocol (and in particular, to the routing protocol)}
  \label{fig:attackEntanglement}
\end{figure}

The most general attack to the routing protocol, pictured in \cref{fig:attackEntanglement}, consists of having two attackers Alice ($\Alice$) and Bob ($\Bob$), located between $\Verif_0$ and $P$, and between $P$ and $\Verif_0$, respectively. Before $t=0$, the attackers agree on a strategy and might prepare an entangled state. After $t=0$, Alice ($\Alice_0$) and Bob ($\Bob_0$) intercept the information sent from their closest verifier, respectively. Due to timing constraints, they are allowed to perform one round of simultaneous communication. After communicating (after $t=1$), Alice ($\Alice_1$) and Bob ($\Bob_1$) answer to their closest verifier, respectively.

Here, we will analyze security in two different models: (i) the \emph{No Pre-shared Entanglement} (No-PE) Model~\cite{Buhrman_2014}, where the adversaries are not allowed to (pre-)share any entanglement before the execution of the protocol, and (ii) the \emph{random oracle model} (ROM), where the attackers are allowed to (pre-)share any amount of entanglement before the execution of the protocol. We formalize the concept of security, given an attack model $\mathcal{M}$, as follows:

\begin{definition}
 The routing protocol is said to be \emph{$\alpha$-sound} in the $\mathcal{M}$ model if, for any attackers acting according to such an attack model, the verifiers \emph{accept} with probability at most $\alpha$.
\end{definition}

The security of a variation of this protocol, the $f$-routing QPV protocol, where the classical information $x$ is split into two bit strings, each sent from each verifier, and the qubit has to be routed according to the outcome of a boolean function $f$ on those bit strings has been studied~\cite{bluhm2022single,asadi2024lineargateboundsnatural,asadi2024ranklowerboundsnonlocal}. The authors of these works showed that the $f$-routing QPV protocol remains secure as long as the attackers pre-share any amount of qubits at the beginning of the protocol that is at most linear in the size of the bit strings. However, unlike other protocols~\cite{Buhrman_2014,TomamichelMonogamyGame2013,allerstorfer2022rolequantumcommunicationloss} the security of the routing QPV protocol in the \emph{No Pre-shared Entanglement} (No-PE) Model~\cite{Buhrman_2014}, where the attackers are not allowed to pre-share entanglement before the execution of the protocol, was never analyzed. The No-PE assumption is necessary to obtain non-trivial bounds, since there is a perfect attack if the attackers pre-share entanglement~\cite{OriginalQPV_Kent2011}. We show security in the No-PE model, providing the tight result, summarized in the following proposition:

\begin{prop}\label{thm:routing_3/4} In the No-PE model, the routing QPV protocol is $\frac{3}{4}$-sound. Moreover, this is optimal. 
\end{prop}
The intuition behind \cref{thm:routing_3/4} relies on the fact that the most general attack can be reduced to a QCG$_2$. Consider the purified version of the routing protocol, which is equivalent to the original version, and where the only difference relies on \textsf{V}$_0$, instead of sending the qubit $\ket\phi$, prepares the state $\ket{\Phi^+}$ and keeps a register for herself and sends the other register to the prover. Then, as seen in \cref{fig:attackEntanglement}, the most general attack to the routing QPV protocol is to place an adversary between \textsf{V}$_0$ and the prover, Alice, and another adversary between the prover and \textsf{V}$_1$, Bob. In the No-PE model, we can simplify it further, as Alice intercepts the qubit sent by \textsf{V}$_0$, applies an arbitrary quantum operation to it, and possibly some ancillary systems she possesses. She keeps a part of it and sends the other to Bob. On the other side, Bob intercepts $x$, copies it and sends the copy to Alice. Since they share no entanglement, any quantum operation that Bob could perform as a function of $x$ can be included in Alice's operation. After one-round of simultaneous communication, Alice and Bob share a tripartite state with \textsf{V}$_0$, and their task is that the party designated by $x$ has to end up with a maximally entangled state with the \textsf{V}$_0$. By \cref{thm:bound_k_parties}, even if Alice and Bob can share any state with the referee (in this case \textsf{V}$_0$), they can succeed with at most probability $\frac{3}{4}$. 

On the other hand, to show optimality, consider the attack where (i) at the beginning of the protocol Alice prepares the 3-qubit state ${\frac{1}{\sqrt{3}}(\ket{\Phi^+}_{A_0A}\ket0_{B}+\ket{\Phi^+}_{A_0B}\ket0_{A})}$, (ii) intercepts $\ket{\phi}$ and performs a Bell measurement on the intercepted state and her register $A_0$, immediately she applies the teleportation corrections to both of her registers $A$ and $B$, (iii) she keeps register $A$ and sends register $B$ to Bob, (iv) in the meantime, Bob intercepts and broadcasts $x$, after receiving the information from their fellow attacker, if $x=0$, Alice sends her register ($A$) to \textsf{V}$_0$, whereas if $x=1$, Bob sends his register ($B$) to \textsf{V}$_1$. This attack has winning a probability of $\frac{3}{4}$. 

An analogous reduction applies when the routing QPV protocol is executed $n$ times in parallel, and therefore, its security can be reduced to the $n$-parallel repetition of QCG$_2$:

\begin{prop}\label{thm:routing_n_parallel} In the No-PE model, the routing QPV protocol executed $n$ times in parallel is ${(\frac{1}{2}+\frac{1}{2\sqrt{2}})^n}$-sound.  
\end{prop}

\ssection{Application to QPV in the random oracle model} Consider the $n$-parallel repetition of the routing QPV protocol but instead of \textsf{V}$_1$ sending $x\in\{0,1\}^n$, \textsf{V}$_0$ and \textsf{V}$_1$ send $r_0, r_1\in\{0,1\}^\ell$, for $\ell\in\mathbb N$, to the prover, respectively. Then, the $x$ used in the rest of the protocol is computed via $x = H(r_0 \oplus r_1)$, for a given hash function $H:\{0,1\}^\ell\rightarrow\{0,1\}^n$. We will denote this variation as $(H,n)$-routing QPV protocol, see \cref{Fig_H_routing} for a schematic representation.

To provide security in the quantum random oracle model against adversaries sharing an arbitrary amount of entanglement, we use some techniques introduced in \cite{Unr14_QuantumPositionVerification}. A quantum random oracle is defined as a fixed function $H\colon \{0,1\}^\ell \to \{0,1\}^n$ that is sampled uniformly at random from the set of functions from $\ell$ bits to $n$ bits\footnote{This can be done by simply sampling a large table $T$ of size $2^\ell$, and outputting $T[r]$ when queried on input $r$. Note that this sampling procedure is not efficient: while having an efficient oracle~\cite{Zha19_HowRecordQuantum} is sometimes required, for instance when working with composable security and computationally bounded distinguishers, or when reducing to problems that are hard only for bounded adversaries, in our case we do not need this additional property since we do a reduction to a problem that is hard even for unbounded adversaries.}. The parties are not given the full description of $H$ directly, but they are given oracle access to $H$, in the sense that they have access to a special gate implementing the unitary $U_H\colon \ket{r}\ket{b} \to \ket{r}\ket{b \xor H(r)}$. We denote the number of queries made by the adversary by $q$. As a proof technique, an oracle can also be reprogrammed, where security of the protocol is shown by first studying a variant where the gate applied by the oracle may change over time. A typical setting is where we change a single entry of the oracle: we denote by $H[r \mapsto x]$ the new oracle that behaves like $H$ except that $H(r) = x$. The chances of distinguishing whether $H$ has been reprogrammed or not can be bounded using~\cite{Unr14_QuantumPositionVerification}, which informally states that if we can distinguish whether the oracle has been reprogrammed or no, then we have queried it on $r$ before it has been reprogrammed (for completeness, see \cref{unr14:reprogrammable}  in the SM). In the following theorem, we show security of the $(H,n)$-routing protocol in the ROM.

\begin{theorem}\label{thm:securityQPVinRO}
  If the (possibly entangled) attackers Alice and Bob perform at most $q$ queries to the (quantum) random oracle $H$, the $(H,n)$-routing QPV protocol is $\epsilon$-sound, with 
  \begin{equation}
    \epsilon=2q2^{-\frac{\ell}{2}}+\left(\frac{1}{2}+\frac{1}{2\sqrt{2}}\right)^n.
  \end{equation}
  In particular, $\eps$ is negligible if $q$ and $\ell$ scale polynomially with $n$.
\end{theorem}
\emph{Proof sketch}. The full proof is in the SM, but the starting idea follows~\cite{Unr14_QuantumPositionVerification}, where we send Bell pairs instead of single qubits in order to make the input state independent of $x$, the output of the oracle. Then we reprogram this oracle \emph{after} adversaries share their state to ensure $x$ is truly random and independent of the state shared by malicious parties at time $t = 1$. Finally, we realize that we can rewrite this into an instance of \cref{thm:optimal2QCGparallel}.

\ssection{Discussion} We have introduced the concept of the $k$-party quantum cloning game and provided the optimal winning probability for any number of parties. The parallel repetition for the two-party version was studied, showing an exponential decay of the optimal winning probability. We applied the above results to show security of the routing QPV protocol in the No-PE model as well as the random oracle model. The tightness of \cref{thm:optimal2QCGparallel} remains an open question, either by showing a strategy attaining the value \eqref{eq:parallel_rep_2_parties}, or if strong parallel repetition holds and actually the optimal value is $\left(\frac{3}{4}\right)^n $ (or neither of them). Closing this gap would imply knowing what is the optimal security for the routing protocol in the No-PE model, and would further tighten the security of the routing protocol in the random-oracle model.

\bibliographystyle{apsrev4-1}
\bibliography{biblio.bib}
\onecolumngrid
\clearpage

\begin{center}
    \textbf{SUPPLEMENTAL MATERIAL}
\end{center}

\noindent{\Large \textbf{Proof of \cref{thm:w_k_parties_EPR}}}\\

\noindent A strategy ${S}$ for the QCG$_k$ is described by a quantum state
$\rho\in \mathcal{S}(\mathcal{H}_R\otimes\mathcal{H}_{P_0E_0}\otimes\ldots\otimes\mathcal{H}_{P_{k-1}E_{k-1}})$, where, for $i\in[k]$, registers $P_i$ are of the same dimension as $\mathcal{H}_R$ and $E_i$ are auxiliary systems of arbitrary dimension that each party possesses, and unitary transformations $U=\{U^x_{P_iE_i}\}_x$,  acting on the registers in the subscripts  (due to the Stinespring dilation of the quantum channels, we restrict our attention to unitary transformations $U^x_{P_iE_i}$ instead of quantum channels $\mathcal{E}^x_{P_iE_i\rightarrow P_i}$). Let $d$ be the dimension of the above (total) Hilbert space, which we denote by $\mathcal H_d$. Then, the winning probability of the QCG$_k$, given the strategy $S$ on a $d$-dimensional Hilbert space, is provided by
\begin{equation}
\begin{split}
    \omega(\text{QCG}_k,S,d)&=\frac{1}{k}\sum_{x\in [k]}\tr{\left(\ketbra{\Phi^+}{\Phi^+}_{RP_x}\otimes \mathbb I_{E_x}\bigotimes_{i\neq x\in[k]}\mathbb I_{P_iE_i}\right)
    \Bigg(\left(\mathbb I_{R}\otimes U^x_{P_iE_i}\otimes\ldots\otimes U^x_{P_iE_i} \right)\rho \left(\mathbb I_{R}\otimes U^{x}_{P_iE_i}\otimes\ldots \otimes U^{x}_{P_iE_i}\right)^{\dagger}\Bigg)}\\&
    =\frac{1}{k}\sum_{x\in [k]}\tr{\left(\mathbb I_{R}\otimes U^{x\dagger}_{P_iE_i}\otimes\ldots \otimes U^{x\dagger}_{P_iE_i}\right)\left(\ketbra{\Phi^+}{\Phi^+}_{RP_x}\otimes \mathbb I_{E_x}\bigotimes_{i\neq x\in[k]}\mathbb I_{P_iE_i}\right)
    \left(\mathbb I_{R}\otimes U^x_{P_iE_i}\otimes\ldots\otimes U^x_{P_iE_i} \right)\rho },
\end{split}
\end{equation}
where in the last equation we used cyclicity of the trace. For a specific choice of  unitary transformations ${U=\{U^x_{P_iE_i}\}_x}$, the optimal winning probability is given by 
\begin{equation}
\begin{split}
    \omega^*(\text{QCG}_k,&U,d)
    =\sup_{\rho\in\mathcal S(\mathcal H_d)}\frac{1}{k}\sum_{x\in [k]}\tr{\left(\mathbb I_{R}\otimes U^{x\dagger}_{P_iE_i}\otimes\ldots \otimes U^{x\dagger}_{P_iE_i}\right)\left(\ketbra{\Phi^+}{\Phi^+}_{RP_x}\otimes \mathbb I_{E_x}\bigotimes_{i\neq x\in[k]}\mathbb I_{P_iE_i}\right)
    \left(\mathbb I_{R}\otimes U^x_{P_iE_i}\otimes\ldots\otimes U^x_{P_iE_i} \right)\rho }
    \\&=
    \frac{1}{k}\norm{\sum_{x\in [k]}\left(\mathbb I_{R}\otimes U^{x\dagger}_{P_iE_i}\otimes\ldots \otimes U^{x\dagger}_{P_iE_i}\right)\left(\ketbra{\Phi^+}{\Phi^+}_{RP_x}\otimes \mathbb I_{E_x}\bigotimes_{i\neq x\in[k]}\mathbb I_{P_iE_i}\right)
    \left(\mathbb I_{R}\otimes U^x_{P_iE_i}\otimes\ldots\otimes U^x_{P_iE_i} \right) }
    \\&=\frac{1}{k}\norm{\sum_{x\in [k]}
    \bigg(\left(\mathbb I_{R}\otimes U^{x\dagger}_{P_xE_x}\right)\left(\ketbra{\Phi^+}{\Phi^+}_{RP_x}\otimes \mathbb I_{E_x}\right)\left(\mathbb I_{R}\otimes U^{x\dagger}_{P_xE_x}\right)\bigg)\bigotimes_{i\neq x\in[k]}U^{x\dagger}_{P_iE_i} \bigotimes_{i\neq x\in[k]}\mathbb I_{P_iE_i} \bigotimes_{i\neq x\in[k]}U^{x}_{P_iE_i}
    }
    \\&=\frac{1}{k}\norm{\sum_{x\in [k]}
    \bigg(\left(\mathbb I_{R}\otimes U^{x\dagger}_{P_xE_x}\right)\left(\ketbra{\Phi^+}{\Phi^+}_{RP_x}\otimes \mathbb I_{E_x}\right)\left(\mathbb I_{R}\otimes U^{x\dagger}_{P_xE_x}\right)\bigg)\bigotimes_{i\neq x\in[k]}U^{x\dagger}_{P_iE_i}U^{x}_{P_iE_i}
    },
        \end{split}
\end{equation}
Notice that, since $\{U^x_{P_iE_i}\}_x$ are unitary matrices, $U^{x\dagger}_{P_iE_i}U^{x}_{P_iE_i}=\mathbb I_{P_iE_i}$, moreover, $\mathbb I_{P_iE_i}=U^{i\dagger}_{P_iE_i}U^{i}_{P_iE_i}$, then we can use $U^{x\dagger}_{P_iE_i}U^{x}_{P_iE_i}=U^{i\dagger}_{P_iE_i}U^{i}_{P_iE_i}$, and therefore
\begin{equation}\label{eq:optimal_independent_on_U}
\begin{split}
\omega^*(\text{QCG}_k,U,d)&
    =\frac{1}{k}\norm{\sum_{x\in [k]}
    \bigg(\left(\mathbb I_{R}\otimes U^{x\dagger}_{P_xE_x}\right)\left(\ketbra{\Phi^+}{\Phi^+}_{RP_x}\otimes \mathbb I_{E_x}\right)\left(\mathbb I_{R}\otimes U^{x\dagger}_{P_xE_x}\right)\bigg)\bigotimes_{i\neq x\in[k]}U^{i\dagger}_{P_iE_i}U^{i}_{P_iE_i}
    }
    \\&=\frac{1}{k}\norm{\sum_{x\in [k]}
    \left(\mathbb I_{R}\bigotimes_{i\neq\in[k]}U^{i\dagger}_{P_iE_i}\right)
    \left(\ketbra{\Phi^+}{\Phi^+}_{RP_x}\otimes \mathbb I_{E_x}\bigotimes_{i\neq x\in[k]}\mathbb I_{P_iE_i}\right)
    \left(\mathbb I_{R}\bigotimes_{i\in[k]}U^{i}_{P_iE_i}\right)}
    \\&=\frac{1}{k}\norm{\left(\mathbb I_{R}\bigotimes_{i\neq\in[k]}U^{i\dagger}_{P_iE_i}\right)
    \left(\sum_{x\in [k]}    
   \ketbra{\Phi^+}{\Phi^+}_{RP_x}\otimes \mathbb I_{E_x}\bigotimes_{i\neq x\in[k]}\mathbb I_{P_iE_i}\right)
    \left(\mathbb I_{R}\bigotimes_{i\in[k]}U^{i}_{P_iE_i}\right)}
    \\&=\frac{1}{k}\norm{\sum_{x\in [k]}    
   \ketbra{\Phi^+}{\Phi^+}_{RP_x}\otimes \mathbb I_{E_x}\bigotimes_{i\neq x\in[k]}\mathbb I_{P_iE_i}}
   \\&=\sup_{\rho\in\mathcal S(\mathcal H_d)}\frac{1}{k}\sum_{x\in [k]}\tr{\left(  
   \ketbra{\Phi^+}{\Phi^+}_{RP_x}\otimes \mathbb I_{E_x}\bigotimes_{i\neq x\in[k]}\mathbb I_{P_iE_i}\right)\rho}
   \\&=\omega^*(\text{QCG}_k,d)
        \end{split}
\end{equation}
where in the fourth equality we used that the  Schatten $\infty$-norm is unitarily invariant, i.e. $\norm{V * W}=\norm{*}$ for unitary matrices $V$ and $W$, and $\omega^*(\text{QCG}_k,d)$ denotes the optimal winning probability if the dimension of the total initial Hilbert space is $d$. Equation~\eqref{eq:optimal_independent_on_U} shows that, given a Hilbert space $\mathcal{H}_R\otimes\mathcal{H}_{P_0E_0}\otimes\ldots\otimes\mathcal{H}_{P_{k-1}E_{k-1}}$, the optimal winning probability can be attained by an optimal quantum state independently of the actions of the players after knowing $x$, i.e.\ the optimal winning probability is independent of $\{U^x_{P_iE_i}\}_x$ and they can apply $\{\mathbb I^x_{P_iE_i}\}_x$ . We are going to see that, actually, the optimal winning probability can be attained by each of the parties possessing a qubit (2-dimensional Hilbert space), i.e.\ by the total Hilbert space being $\mathcal{H}_{2^k}=\bigotimes_{i\in[k]}\mathbb C^2$. From \eqref{eq:optimal_independent_on_U},
\begin{equation}\label{eq:optimal_by_qubits}
\begin{split}
    \omega^*(\text{QCG}_k)&=\sup_{d\in\mathbb N}\omega^*(\text{QCG}_k,d)=\sup_{d\in\mathbb N}\frac{1}{k}\norm{\left(\sum_{x\in [k]}    
   \ketbra{\Phi^+}{\Phi^+}_{RP_x}\otimes \mathbb{I}_{P_0\ldots \slashed{P}_{x} \ldots P_{k-1}}\right)\bigotimes_{i\in[k]}\mathbb I_{E_i}}   \\&
   =\sup_{d\in\mathbb N}\frac{1}{k}\norm{\sum_{x\in [k]}    
   \ketbra{\Phi^+}{\Phi^+}_{RP_x}\otimes \mathbb{I}_{P_0\ldots \slashed{P}_{x} \ldots P_{k-1}}}\norm{\bigotimes_{i\in[k]}\mathbb I_{E_i}} = \sup_{d\in\mathbb N} \frac{1}{k}\norm{\sum_{x\in [k]}    
   \ketbra{\Phi^+}{\Phi^+}_{RP_x}\otimes \mathbb{I}_{P_0\ldots \slashed{P}_{x} \ldots P_{k-1}}}\\&=
   \frac{1}{k}\norm{\sum_{x\in [k]}    
   \ketbra{\Phi^+}{\Phi^+}_{RP_x}\otimes \mathbb{I}_{P_0\ldots \slashed{P}_{x} \ldots P_{k-1}}} \sup_{d\in\mathbb N}= \frac{1}{k}\norm{\sum_{x\in [k]}    
   \ketbra{\Phi^+}{\Phi^+}_{RP_x}\otimes \mathbb{I}_{P_0\ldots \slashed{P}_{x} \ldots P_{k-1}}} \\&=\sup_{\rho\in\mathcal S\left(\mathcal{H}_{2^k}\right)} \frac{1}{k}\sum_{x\in [k]} \tr{\left(   
   \ketbra{\Phi^+}{\Phi^+}_{RP_x}\otimes \mathbb{I}_{P_0\ldots \slashed{P}_{x} \ldots P_{k-1}}\right)\rho}, 
    \end{split}
\end{equation}
where, in the arguments of the supremums, the dependence on $d$ is implicit in the auxiliary spaces, which, together with the registers $P_i$ and $V$, fully describe the total Hilbert space, and thus its dimension. 

In order to provide the explicit value for the optimal winning probability, we have that, from \eqref{eq:optimal_by_qubits},
\begin{equation}
     \omega^*(\text{QCG}_k)=\frac{1}{k}\norm{\sum_{x\in [k]}    
   \ketbra{\Phi^+}{\Phi^+}_{RP_x}\otimes\mathbb{I}_{P_0\ldots \slashed{P}_{x} \ldots P_{k-1}}}=\frac{1}{2}+\frac{1}{2k}, 
\end{equation}
where the last equation is obtained by direct computation.\\

\noindent{\Large \textbf{Proof of \cref{thm:bound_k_parties}}}\\

\noindent The result follows from the proof of \cref{thm:w_k_parties_EPR} by repeating the same steps, replacing $\ket{\Phi^+}_{VP_x}$ by $\ket{\Psi}_{VP_x}$, and from \eqref{eq:optimal_by_qubits}, we obtain
\begin{equation}
    \omega^*(\Psi\text{\normalfont{-QCG}}_k)=\frac{1}{k}\norm{\sum_{x\in [k]}\ketbra{\Psi}{\Psi}_{VP_x}\otimes \mathbb{I}_{P_0\ldots \slashed{P}_{x} \ldots P_{k-1}}}.
\end{equation}

\noindent{\Large \textbf{Proof of \cref{thm:routing_n_parallel}}}\\

\noindent We show the proof of \cref{thm:routing_n_parallel}. For that, we need the following definition and lemmas. 
\begin{definition} Let $n\in\mathbb N$. Two permutations $\pi,\pi':[n]\rightarrow[n]$ are said to be \emph{orthogonal} if $\pi(i)\neq\pi'(i)$ for all $i\in[n]$. 
\end{definition}

\begin{lemma} \label{lem:sum_projectors}\emph{(Lemma 2 in \cite{TomamichelMonogamyGame2013})}
    Let $\Pi^1,\ldots,\Pi^n$ be projectors acting on a Hilbert space $\mathcal H$. Let $\{\pi_k\}_{k\in[n]}$ be a set of mutually orthogonal permutations. Then, 
    \begin{equation}
        \bigg\|\sum_{i\in[n]}\Pi^i \bigg\|\leq \sum_{k\in[n]}\max_{i\in[n]}\big\| \Pi^i\Pi^{\pi_k(i)}\big\|.
    \end{equation}
\end{lemma}

\begin{remark}
    There always exist a set of $n$ permutations of $[n]$ that are mutually orthogonal, an example is the $n$ cyclic shifts.  
\end{remark}

\begin{lemma}\label{lem:norm_product}(Lemma 1 in \cite{TomamichelMonogamyGame2013})
    Let $A,B,L\in\mathcal{B}(\mathcal{H})$ such that $AA^\dagger\succeq B^\dagger B$. Then it holds that $\norm{AL}\geq\norm{BL}$. 
\end{lemma}

Now, we are in position to prove \cref{thm:routing_n_parallel}. A strategy ${S}_n$ for the $n$-parallel repetition of QCG$_2$ is described by a quantum state
$\rho\in \mathcal{S}(\mathcal{H}_R\otimes\mathcal{H}_{A_0\ldots A_{n-1}E_A}\otimes\mathcal{H}_{B_0\ldots B_{n-1}E_B})$, where, for $i\in[n]$, registers $A_i$ and $B_i$ are of the same dimension as $\mathcal{H}_R$ and $E_A$ and $E_B$ are auxiliary systems of arbitrary dimension that each party possess, and unitary transformations $\{U^x_{A_0\ldots A_{n-1}E_A}\}_x$ and $\{V^x_{B_0\ldots B_{n-1}E_B}\}_x$, acting on the registers in the subscripts (due to the Stinespring dilation of the quantum channels, we restrict our attention to unitary transformations). For $x=x_0\ldots x_{n-1}\in\{0,1\}^n$, let $Q_{x_i}=A_i$ if $x_i=0$ and $Q_{x_i}=B_i$ if $x_i=1$, and we use the shorthand notation $R=R_0\ldots R_{n-1}$, $A=A_0\ldots A_{n-1}$ and $B=B_0\ldots B_{n-1}$. Then, the winning probability of this game, given the strategy $S_n$, is provided by 

\begin{equation} 
\begin{split}
    \omega(\normalfont{\text{QCG}}_2^{\times n},S_n)&=\frac{1}{2^n}\sum_{x\in\{0,1\}^n}\tr{ \left(\bigg(\bigotimes_{i\in[n]}\ketbra{\Phi^+}{\Phi^+}_{R_iQ_{x_i}}\otimes\mathbb I_{Q_{1-x_i}}\bigg)\otimes \mathbb I_{E_AE_B}\right)(\mathbb I_{R}\otimes U^x_{AE_A}\otimes V^x_{BE_B})\rho(\mathbb I_{R}\otimes U^x_{AE_A}\otimes V^x_{BE_B})^{\dagger}}\\
    &=\frac{1}{2^n}\sum_{x\in\{0,1\}^n}\tr{(\mathbb I_{R}\otimes U^{x\dagger}_{AE_A}\otimes V^{x\dagger}_{BE_B}) \left(\bigg(\bigotimes_{i\in[n]}\ketbra{\Phi^+}{\Phi^+}_{R_iQ_{x_i}}\otimes\mathbb I_{Q_{1-x_i}}\bigg)\otimes \mathbb I_{E_AE_B}\right)(\mathbb I_{R}\otimes U^x_{AE_A}\otimes V^x_{BE_B})\rho}\\
    &\leq\frac{1}{2^n}\norm{\sum_{x\in\{0,1\}^n}(\mathbb I_{R}\otimes U^{x\dagger}_{AE_A}\otimes V^{x\dagger}_{BE_B}) \left(\bigg(\bigotimes_{i\in[n]}\ketbra{\Phi^+}{\Phi^+}_{R_iQ_{x_i}}\otimes\mathbb I_{Q_{1-x_i}}\bigg)\otimes \mathbb I_{E_AE_B}\right)(\mathbb I_{R}\otimes U^x_{AE_A}\otimes V^x_{BE_B})}.
\end{split}
\end{equation}
Denote
\begin{equation}
    M^x:=(\mathbb I_{R}\otimes U^{x\dagger}_{AE_A}\otimes V^{x\dagger}_{BE_B}) \left(\bigg(\bigotimes_{i\in[n]}\ketbra{\Phi^+}{\Phi^+}_{R_iQ_{x_i}}\otimes\mathbb I_{Q_{1-x_i}}\bigg)\otimes \mathbb I_{E_AE_B}\right)(\mathbb I_{R}\otimes U^x_{AE_A}\otimes V^x_{BE_B}),
\end{equation}
then, 
\begin{equation}
    \omega(\normalfont{\text{QCG}}_2^{\times n},S_n)\leq \frac{1}{2^n}\norm{\sum_{x\in\{0,1\}^n}M^x}\leq\frac{1}{2^n}\sum_{k\in[n]}\max_{x,x'}\norm{M^xM^{x'}},
\end{equation}
where we used Lemma~\ref{lem:sum_projectors}, and $x'=\pi_k(x)$, for $\{\pi_k\}_k$ being a set of mutually orthogonal permutations. Fix $x$ and $x'$, and let $\mathcal T$ be the set of indices where $x$ and $x'$ differ, i.e.~$\mathcal{T}=\{i\mid x_i\neq x'_i\}$, and let $t=\abs{\mathcal{T}}$. Let $\mathcal{T}_A=\{i\in \mathcal{T}\mid x_i=0\}$, and $t_A:=\abs{\mathcal{T}_A}$, then we have that
\begin{equation}
    \begin{split}
        M^x \preceq M^x_A&:=(\mathbb I_{R}\otimes U^{x\dagger}_{AE_A}\otimes V^{x\dagger}_{BE_B})\left(\bigg(\bigotimes_{i\in \mathcal{T}_A}\ketbra{\Phi^+}{\Phi^+}_{R_iQ_{x_i}}\otimes\mathbb I_{Q_{1-x_i}}\bigg)\otimes\bigg(\bigotimes_{i\in [n]\setminus\mathcal{T}_A}\mathbb I_{R_iQ_{x_i}Q_{1-x_i}}\bigg)\otimes \mathbb I_{E_AE_B}\right)(\mathbb I_{R}\otimes U^x_{AE_A}\otimes V^x_{BE_B})\\
        &=(\mathbb I_{R}\otimes U^{x\dagger}_{AE_A}\otimes V^{x \dagger}_{BE_B})\left(\bigg(\bigotimes_{i\in \mathcal{T}_A}\ketbra{\Phi^+}{\Phi^+}_{R_iA_{i}}\bigotimes_{i\in [n]\setminus\mathcal{T}_A}\mathbb{I}_{R_iA_{i}E_A}\right)\otimes \mathbb I_{BE_B} \bigg)(\mathbb I_{R}\otimes U^x_{AE_A}\otimes V^{x}_{BE_B})\\
        &=(\mathbb I_{R}\otimes U^{x\dagger}_{AE_A}\otimes V^{x '\dagger}_{BE_B})\left(\bigg(\bigotimes_{i\in \mathcal{T}_A}\ketbra{\Phi^+}{\Phi^+}_{R_iA_{i}}\bigotimes_{i\in [n]\setminus\mathcal{T}_A}\mathbb{I}_{R_iA_{i}E_A}\right)\otimes \mathbb I_{BE_B} \bigg)(\mathbb I_{R}\otimes U^x_{AE_A}\otimes V^{x'}_{BE_B}),
            \end{split}
\end{equation}
where in the last equality we used that $V^{x \dagger}_{BE_B}V^{x \dagger}_{BE_B}=\mathbb I_{BE_B}=V^{x' \dagger}_{BE_B}V^{x'}_{BE_B}$. Similarly, 

\begin{equation}
    \begin{split}
        M^{x'} \preceq M^{x'}_B&:=(\mathbb I_{R}\otimes U^{x'\dagger}_{AE_A}\otimes V^{x'\dagger}_{BE_B})\left(\bigg(\bigotimes_{i\in \mathcal{T}_A}\ketbra{\Phi^+}{\Phi^+}_{R_iQ_{x'_i}}\otimes\mathbb I_{Q_{1-x'_i}}\bigg)\otimes\bigg(\bigotimes_{i\in [n]\setminus\mathcal{T}_A}\mathbb I_{R_iQ_{x'_i}Q_{1-x'_i}}\bigg)\otimes \mathbb I_{E_AE_B}\right)(\mathbb I_{R}\otimes U^{x'}_{AE_A}\otimes V^{x'}_{BE_B})\\
        &=(\mathbb I_{R}\otimes U^{x'\dagger}_{AE_A}\otimes V^{x' \dagger}_{BE_B})\left(\bigg(\bigotimes_{i\in \mathcal{T}_A}\ketbra{\Phi^+}{\Phi^+}_{R_iB_{i}}\bigotimes_{i\in [n]\setminus\mathcal{T}_A}\mathbb{I}_{R_iB_{i}E_B}\right)\otimes \mathbb I_{AE_A} \bigg)(\mathbb I_{R}\otimes U^{x'}_{AE_A}\otimes V^{x'}_{BE_B})\\
        &=(\mathbb I_{R}\otimes U^{x\dagger}_{AE_A}\otimes V^{x' \dagger}_{BE_B})\left(\bigg(\bigotimes_{i\in \mathcal{T}_A}\ketbra{\Phi^+}{\Phi^+}_{R_iB_{i}}\bigotimes_{i\in [n]\setminus\mathcal{T}_A}\mathbb{I}_{R_iB_{i}E_B}\right)\otimes \mathbb I_{AE_A} \bigg)(\mathbb I_{R}\otimes U^x_{AE_A}\otimes V^{x'}_{BE_B}),
    \end{split}
\end{equation}
By Lemma~\ref{lem:norm_product}, 
\begin{equation}\label{eq:MxMx<=...}
    \norm{M^xM^{x'}}\leq\norm{M^x_AM^{x'}_B},
\end{equation}
then
\begin{equation}
\begin{split}
    M^x_AM^{x'}_B=&(\mathbb I_{R}\otimes U^{x\dagger}_{AE_A}\otimes V^{x '\dagger}_{BE_B})\left(\bigg(\bigotimes_{i\in \mathcal{T}_A}\ketbra{\Phi^+}{\Phi^+}_{R_iA_{i}}\bigotimes_{i\in [n]\setminus\mathcal{T}_A}\mathbb{I}_{R_iA_{i}E_A}\right)\otimes \mathbb I_{BE_B} \bigg)(\mathbb I_{R}\otimes U^x_{AE_A}\otimes V^{x'}_{BE_B})\\
    &\cdot (\mathbb I_{R}\otimes U^{x\dagger}_{AE_A}\otimes V^{x' \dagger}_{BE_B})\left(\bigg(\bigotimes_{i\in \mathcal{T}_A}\ketbra{\Phi^+}{\Phi^+}_{R_iB_{i}}\bigotimes_{i\in [n]\setminus\mathcal{T}_A}\mathbb{I}_{R_iB_{i}E_B}\right)\otimes \mathbb I_{AE_A} \bigg)(\mathbb I_{R}\otimes U^x_{AE_A}\otimes V^{x'}_{BE_B}).
    \end{split}
\end{equation}

We have that $(\mathbb I_{R}\otimes U^x_{AE_A}\otimes V^{x'}_{BE_B})(\mathbb I_{R}\otimes U^{x\dagger}_{AE_A}\otimes V^{x' \dagger}_{BE_B})=\mathbb{I}_{RAEABE_B}$, and, since the Schatten $\infty$-norm is unitarily invariant, 
\begin{equation}\label{eq:normMxAMxB}
\begin{split}
     \norm{ M^x_AM^{x'}_B}&=\norm{\left(\bigotimes_{i\in \mathcal{T}_A}\ketbra{\Phi^+}{\Phi^+}_{R_iA_{i}}\bigotimes_{i\in [n]\setminus\mathcal{T}_A}\mathbb{I}_{R_iA_{i}E_A}\otimes \mathbb I_{BE_B} \right)\left(\bigotimes_{i\in \mathcal{T}_A}\ketbra{\Phi^+}{\Phi^+}_{R_iB_{i}}\bigotimes_{i\in [n]\setminus\mathcal{T}_A}\mathbb{I}_{R_iB_{i}E_B}\otimes \mathbb I_{AE_A} \right)}\\
     &=\norm{\left(\bigotimes_{i\in \mathcal{T}_A}\big(\ketbra{\Phi^+}{\Phi^+}_{R_iA_{i}}\otimes\mathbb I_{B_i}\big)\big(\ketbra{\Phi^+}{\Phi^+}_{R_iB_{i}}\otimes\mathbb I_{A_i}\big)\right)\bigotimes_{i\in [n]\setminus\mathcal{T}_A}\mathbb I_{R_iA_iB_i}\otimes\mathbb{I}_{E_AE_B}}
     \\&=\norm{\bigotimes_{i\in \mathcal{T}_A}\big(\ketbra{\Phi^+}{\Phi^+}_{R_iA_{i}}\otimes\mathbb I_{B_i}\big)\big(\ketbra{\Phi^+}{\Phi^+}_{R_iB_{i}}\otimes\mathbb I_{A_i}\big)}\norm{\bigotimes_{i\in [n]\setminus\mathcal{T}_A}\mathbb I_{R_iA_iB_i}\otimes\mathbb{I}_{E_AE_B}}\\
     &=\prod_{i\in\mathcal{T}_A}\big(\ketbra{\Phi^+}{\Phi^+}_{R_iA_{i}}\otimes\mathbb I_{B_i}\big)\big(\ketbra{\Phi^+}{\Phi^+}_{R_iB_{i}}\otimes\mathbb I_{A_i}\big)
     \\&=2^{-t_A},
\end{split}
\end{equation}
where we used that, for every $i$,
\begin{equation}
    \norm{\big(\ketbra{\Phi^+}{\Phi^+}_{R_iA_{i}}\otimes\mathbb I_{B_i}\big)\big(\ketbra{\Phi^+}{\Phi^+}_{R_iB_{i}}\otimes\mathbb I_{A_i}\big)}=2^{-1}.
\end{equation}
Without loss of generality, assume $t_A\geq t/2$, then, combining \eqref{eq:MxMx<=...} and \eqref{eq:normMxAMxB}, we have that 
\begin{equation}
    \norm{M^xM^{x'}}\leq \norm{M^{x}_AM^{x'}_B}\leq 2^{-\frac{t}{2}}.
\end{equation}

In order to apply the bound in Lemma~\ref{lem:norm_product}, consider the set of permutations given by $\pi_k(x)=x\oplus k$, where $x, k\in \{0,1\}^{n}$ (they are such that they have the same Hamming distance). There are $\binom{n}{i}$ permutations with Hamming distance $i$. Then, we have
\begin{equation}
    \omega(\normalfont{\text{QCG}}_2^{\times n},S_n)\leq \sum_{k\in[n]}\max_{x,x'}\norm{M^xM^{x'}}\leq \frac{1}{2^n}\sum_{t=0}^n\binom{n}{t}2^{-\frac{t}{2}}=\left(\frac{1}{2}+\frac{1}{2\sqrt{2}}\right)^n.
\end{equation}

\noindent{\Large \textbf{Proof of \cref{thm:securityQPVinRO}}}\\

\noindent In the proof of \cref{thm:securityQPVinRO}, we will rely on this lemma by Unruh:

\begin{lemma}[{\cite[Lemma~3]{Unr14_QuantumPositionVerification}}]\label{unr14:reprogrammable}
  Let $(\cA_1, \cA_2)$ be oracle algorithms sharing state between invocations that perform at most $q$ queries to $H$. Let $C$ be an oracle algorithm that on input $(j,r)$ does the following: Run $\cA_1^H(r)$ until the $j$-th query to $H$, then measure the argument of that query in the computational basis, and output the measurement outcome (or $\bot$ if no $j$-th query occurs). Let:
  \begin{align}
    P_\cA^1 &\eqdef \pr*[H \sample (\{0,1\}^\ell \to \{0,1\}^n)\\r \sample \{0, 1\}^l, \cA_1^H(r)\\b' \gets \cA_2^H(r, H(r))]{b' = \accept},\\
    P_\cA^2 &\eqdef \pr*[H \sample (\{0,1\}^\ell \to \{0,1\}^n)\\r \sample \{0, 1\}^l, x \sample \{0,1\}^n\\H' \eqdef H[r \mapsto x], \cA_1^H(r)\\b' \gets \cA_2^{H'}(r, x)]{b' = \accept},\\
    P_C &\eqdef \pr*[H \sample (\{0,1\}^\ell \to \{0,1\}^n)\\r \sample \{0, 1\}^l, j \sample \{1,\dots,q\}\\x' \gets C^H(j,x)]{x = x'}.
  \end{align}
  Then, $|P_\cA^1 - P_\cA^2| \leq 2q\sqrt{P_C}$.
\end{lemma}

  Then, to prove this theorem, we must show that the probability that the verifiers accept in a malicious run of the protocol is lower bounded by $\eps$, i.e., if we denote by $\Verif_0 \interacts \Alice \interacts \Bob \interacts \Verif_1$ the output of the verifiers ($\accept$ or $\reject$) at the end of a protocol involving a malicious Alice $\Alice$ and a malicous Bob $\Bob$, we want to show that
  \begin{align}
    \pr{\Verif_0 \interacts \Alice \interacts \Bob \interacts \Verif_1 = \accept} \leq \eps.  
  \end{align}
  We prove this by defining a series of games, where the probability of accepting each game is close to the probability of accepting the next game. By ensuring that the first game corresponds to the real protocol, and that the probability of the last game can easily be computed, we can bound $\eps$ by transitivity. 

  \textbf{\game{1}}: this game is defined as the real protocol, i.e.\ $\game{1} \eqdef \Verif_0 \interacts \Alice \interacts \Bob \interacts \Verif_1$. Therefore, we trivially have:
  \begin{align}
    \pr{\Verif_0 \interacts \Alice \interacts \Bob \interacts \Verif_1 = \accept} = \pr{\game{1} = \accept}.
  \end{align}

  \textbf{\game{2}}: is like $\game{1}$, except that each $\ket{\phi_i}$ is replaced with one half of a Bell pair. Similarly, instead of projecting on $\ket{\phi_i}$, the verifier will do a Bell measurement between the state sent by the prover and its corresponding half of Bell pair, accepting only if the outcome is $(0,0)$. This trick is often used in literature, hence we skip the computations. 
  Hence
  \begin{align}
    \pr{\game{1} = \accept} = \pr{\game{2} = \accept}.
  \end{align}
  \textbf{\game{3}}: is like $\game{2}$, except that at time $t = 1$, one samples the random bit string $x \sample \{0, 1\}^n$, and reprogram the oracle to implement $H' \eqdef H[r_0 \xor r_1 \mapsto x]$ (i.e., $A_1$ and $B_1$ will have oracle access to $H'$ instead of $H$). Note that the simulators will use this value of $x = H'(r_0 \xor r_1)$ instead of $H(r_0 \xor r_1)$ to perform the verification at the end. Intuitively, the only way to distinguish this game from the previous game is if the adversary managed to query $H(r_0 \xor r_1)$ before $t = 0$ and after, but this is highly unlikely since neither $A_0$ nor $B_0$ know both $r_0$ and $r_1$ (and remember that they cannot query the oracle more than $q$ times, so they cannot just evaluate the oracle on all inputs). This intuition is formalized thanks to \cref{unr14:reprogrammable}: if we define $\cA_1(r)$ as the execution of the protocol in \game{2} until $t=1$ (which is the same as in \game{3}), except that $r_2$ is chosen as $r_2 \eqdef r_1 \xor r$, and $\cA_2(r,x)$ as the execution of the protocol after time $t = 1$, we can remark that (using notations from \cref{unr14:reprogrammable}):
  \begin{align}
    P_\cA^1 = \pr{\game{2} = \accept}.
  \end{align}
  since sampling $(r_1,r_2)$ uniformly at random is strictly equivalent to sampling $(r_1,r)$ randomly and then defining $r_2 \eqdef r \xor r_1$. Similarly, we also have:
  \begin{align}
    P_\cA^2 = \pr{\game{3} = \accept}\label{eq:pcagame3}.
  \end{align}
  The remaining part is to bound $P_C$. To compute $P_C$, we need to bound the probability of querying $H(r)$ during the first part of the protocol on $j$-th query. But when $\Alice_0$ does this query, we know that it must be independent of $r$ since all inputs of $\Alice_0$ are independent of $r$ (if not, we could break non-signaling). Similarly, queries made by $\Bob_0$ are independent of $r$: the exact same argument does not hold since $r_2 = r_1 \xor r$ does depend on $r$\dots{} but this is only a very superficial dependency, since we could have exactly the same probability distributions of $r_1$ and $r_2$ by sampling instead $r_2$ randomly and $r_1 = r_2 \xor r$, making $r_2$ independent of $r$ now. Hence, the $j$-th query is independent of $r$, so the best probability of it being equal to $r$ is lower bounded by $P_C \leq \frac{1}{2^\ell}$. Hence, using \cref{unr14:reprogrammable}, we have:
  \begin{align}
    &|\pr{\game{2} = \accept} - \pr{\game{3} = \accept}|\\
    &\eqeq{eq:pcagame3} |P_\cA^1 - P_\cA^2| \leqref{unr14:reprogrammable} 2q\sqrt{P_C} \leq 2q2^{-\ell/2}.
  \end{align}

  \textbf{\game{4}}: now, we can realize that all operations in $\game{3}$ until $t=1$ is independent of $x$. So let us call $\ket{\psi}_{R P_A P_B}$ the (purification) of the state owned by the verifier (consisting in a list of qubits part of a shared Bell pair), Alice and Bob (we also include in $P_B$ the message sent by $\Alice_0$ to $\Bob_1$, and similarly in $P_A$ the message sent by $\Bob_0$ to $\Alice_1$). Additionally, we also include in $\ket{\psi}$ the (exponentially large) definition of $H$, $r_0$ and $r_1$ in both registers $P_A$ and $P_B$, one copy for each party. Then, we define the referee operation $R$ as the identity, $P_A$ as the map that runs $\Alice_1$, simulating the query to $H'$ using the table $H$, $r_0$ and $r_1$ that are part of $\ket{\psi}$, and $x$ that is given as an input to $P_A$, and we define similarly $P_B$ simulating $\Bob_1$. We define then $\game{4}$ as the game $\text{QCG}_2^{\times n}$, i.e.\ the parallel repetition of the $2$-party quantum cloning game (\cref{def:kpartyQCG}), involving the shared state $\ket{\psi}$, the referee $R$, and the two parties $P_A$ and $P_B$. This game is exactly like $\game{3}$ as we simulate exactly the same process, just grouping differently the various circuits involved. Hence, $\pr{\game{4} = \accept} = \pr{\game{3} = \accept}$. But using \cref{thm:optimal2QCGparallel}, we have:
\begin{align}
  \pr{\game{4}}
  = \omega^*(\emph{\normalfont{QCG}}_2^{\times n})
  \leqref{thm:optimal2QCGparallel}\left(\frac{1}{2}+\frac{1}{2\sqrt{2}}\right)^n.
\end{align}
Hence, we can combine all the above equations to obtain:
\begin{align}
  \pr{\Verif_0 \interacts \Alice \interacts \Bob \interacts \Verif_1 = \accept}
  = \pr{\game{1} = \accept} 
   \leq \left(\frac{1}{2} + \frac{1}{2\sqrt{2}}\right)^n +  2q2^{-\ell/2},
\end{align}
concluding the proof.

\end{document}